\documentclass[aps,prl,twocolumn,showpacs,superscriptaddress]{revtex4}

\RequirePackage{lineno}

\usepackage{hyperref}
\usepackage{graphicx}
\usepackage[utf8]{inputenc}
\DeclareGraphicsExtensions{.png,.jpg,.eps,.pdf}

\usepackage{xcolor}
\usepackage{amsmath}

\begin{document}

\title{Ising and XY paramagnons in two-dimensional 2H-NbSe$_2$}

\author{A. T. Costa}
\affiliation{QuantaLab, International Iberian Nanotechnology Laboratory (INL), 
Av. Mestre Jos\'e Veiga, 4715-330 Braga, Portugal}
\author{M. Costa} 
\affiliation{Instituto de F\'isica, Universidade Federal Fluminense, 
24210-346, Niter\'oi, RJ, Brazil}

\author{J. Fern\'andez-Rossier}
\altaffiliation[On leave from ]{Departamento de F\'{\i}sica Aplicada, Universidad de Alicante, 03690,  Sant Vicent del Raspeig, Spain }
\affiliation{QuantaLab, International Iberian Nanotechnology Laboratory (INL), 
Av. Mestre Jos\'e Veiga, 4715-330 Braga, Portugal}

\date{\today}

\begin{abstract}

Paramagnons are the collective modes that govern the spin response of nearly 
magnetic conductors. In some cases they mediate electron pairing leading 
to superconductivity. This scenario may occur in 2H-NbSe$_2$ monolayers, that feature
spin-valley coupling on account of spin-orbit interactions and their lack of inversion symmetry.
Here we explore spin anisotropy of paramagnons both for non-centrosymmetric  Kane-Mele-Hubbard 
models  for  2H-NbSe$_2$ monolayers described with  a DFT-derived tight-binding model.
In the infinite wavelength limit we find spatially uniform  paramagnons with energies 
around $1$~meV that feature a colossal off-plane uniaxial magnetic anisotropy, 
with quenched transversal spin response.
At finite wave vectors, longitudinal and transverse channels 
reverse roles: XY fluctuations dominate within a significant portion of the 
Brillouin zone. Our results show that 2H-NbSe$_2$ is close to a Coulomb-driven
in-plane (XY) spin density wave instability.
\end{abstract}

\maketitle

A nearly magnetic conductor is a material on the brink of a quantum phase transition 
to a magnetically ordered state.  The transition is controlled by the Stoner parameter, 
defined as the product of the atomic Coulomb repulsion $U$ and the density of states at the Fermi
energy, $\rho_0$. As it happens in conventional phase transitions, fluctuations are enhanced 
due to proximity to the critical point. In the case of nearly ferromagnetic conductors, 
spin fluctuations are enhanced when $\rho_0 U \simeq 1$, leading to the emergence of 
paramagnons, prominent features in the low energy spectra, that anticipate the formation 
of magnon resonances at the other side of the transition.  The formation of paramagnons occurs for all
magnetic instabilities, either ferromagnetic, antiferromagnetici, or spin-density wave,  
and lead to diverging magnetic responses at specific wave-vectors  that characterize the ordered  phase at the other side of the transition.~\cite{BookMoriya}.

Interaction of paramagnons with quasiparticles lead to observable effects, such as the 
renormalizaton of the quasiparticle effective mass~\cite{PhysRevLett.17.750} and a resulting enhancement
of the electrical resistivity~\cite{PhysRev.165.837} and electronic specific 
heat~\cite{PhysRevLett.17.750,BerkSchrieffer1966}. Ferromagnetic spin fluctuations can also result in p-wave 
triplet pairing\cite{rice1995,monthoux2007}, that could lead to the coexistence of triplet 
SC and FM, or the emergence of SC order in the vicinity of a FM phase transition. 
The interplay between superconductivity and ferromagnetic spin fluctuations has been 
explored in materials like Pd\cite{fay1977}, ZnZr$_2$\cite{pfleiderer2001}, liquid $^3$He, 
twisted bilayer graphene\cite{Cao2018,PhysRevLett.121.087001,You2019}, ABC graphene trilayer\cite{AFYoung2021,Levitov2021}, UTe$_2$\cite{ran2019}, 
and 2H-NbSe$_2$\cite{wickramaratne2020,wan2021,Hamill2021}. Whereas most of these materials are 
centro-symmetric and  spin-orbit coupling (SOC) has a minor impact and is customarily neglected, 
the case of 2H-NbSe$_2$ monolayers is very different.  

\begin{figure}
 \centering
    \includegraphics[width=\columnwidth]{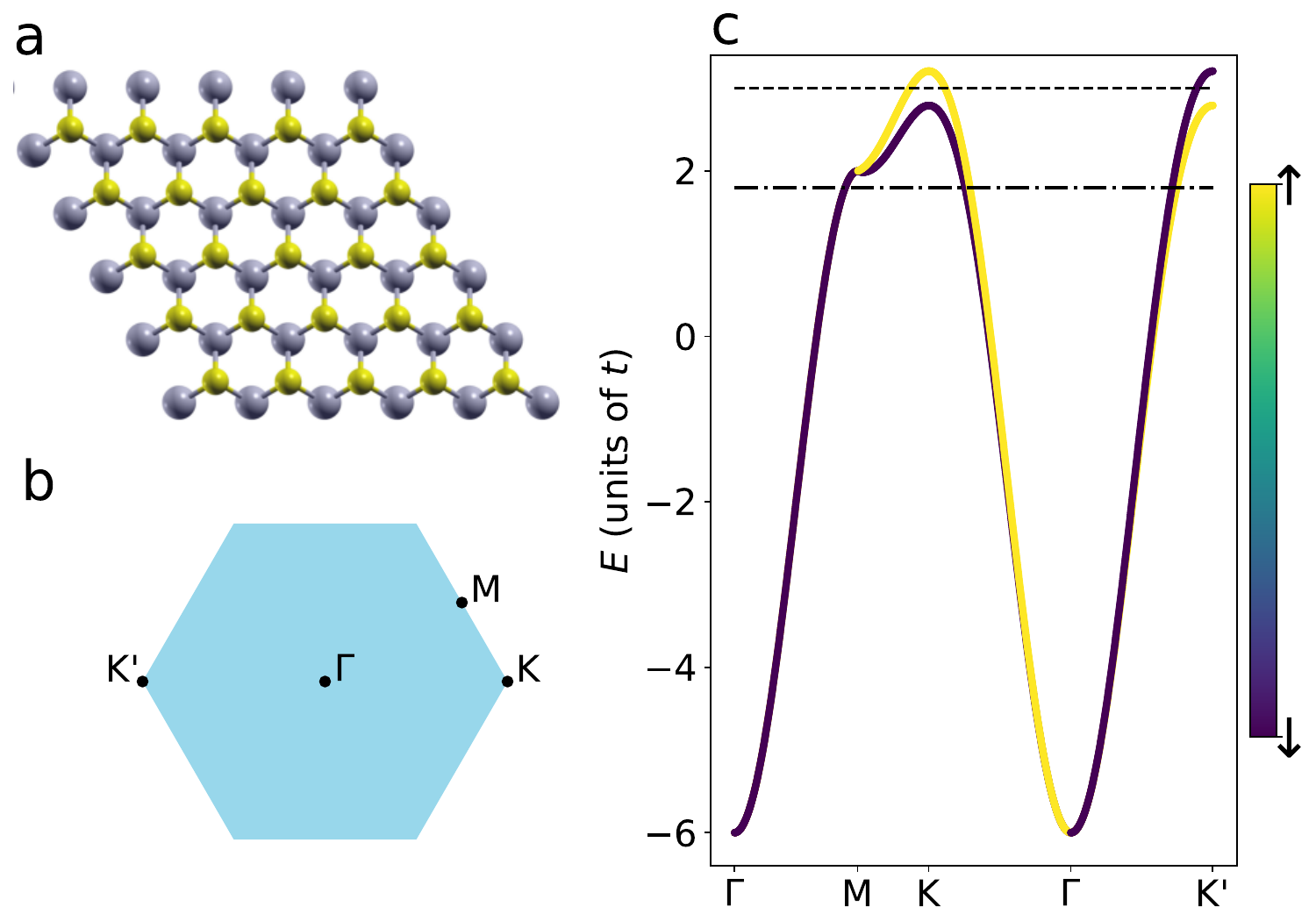}
\caption
{a) NbSe$_{2}$ honeycomb lattice (HCL) with broken inversion symmetry. 
b) Brillouin zone of the HCL displaying the relevant high-symmetry
points. c) Band structure of the extended Kane-Mele model with 
$\Delta\rightarrow\infty$ and finite SOC ($t_{KM}=0.04t$). 
The two horizontal lines mark the values of $E_F$ used in the
calculations of the spin fluctuation spectra displayed in Fig.~\ref{fig2}.
Dashed line: $E_F=3t$, dot-dashed line: $E_F=1.8t$. }
\label{fig1}
\end{figure}

Spin-orbit interaction has a dramatic effect on the energy bands of two-dimensional 
2H-NbSe$_2$ and related transition metal dichalcogenide (TMD) monolayers\cite{Xiao12,Kosmider13b}. 
The lack of inversion symmetry leads to a momentum-dependent spin splitting of the energy 
bands. The splitting is large, on account of the strong SOC
of the transition metal. As a result, Kramers doublets have their momenta at opposite points 
in the Brillouin zone (see Fig.~\ref{fig1}). For the states at the corner points of the BZ, the 
so-called valleys, this phenomenon is the celebrated spin-valley coupling, that leads to a 
peculiar band structure, with two pockets that feature complete and opposite spin 
polarizations.

In this paper, we study spin fluctuations in spin-valley coupled systems.  
From inspection of their energy bands we can expect a very anisotropic 
spin response. When the Fermi energy lies in the half-metallic pockets 
at the top of the valence band (see Fig.~\ref{fig1}), spin-flip fluctuations 
are gapped for $q=0$, in contrast with longitudinal spin-conserving fluctuations.  
This effect also occurs when the Fermi surface is no longer at the valleys, but 
still in the spin-split region. The first case is relevant for 2H-MoS$_2$, for 
which a doping induced ferromagnetic transition has been reported\cite{roch2019}, 
and other semiconducting TMDs. The second case is relevant for 2H-NbSe$_2$.  
To study this phenomenon, we compute spin fluctuations using the Random Phase 
Approximation (RPA) for two types of Hamiltonians. First, we consider the 
Kane-Mele-Hubbard model\cite{rachel2010,soriano2010,fukaya2016,wu2019} with 
a sublattice potential term that breaks inversion symmetry, leading to spin-valley 
coupled bands. Second, we consider a multi-orbital effective Hamiltonian 
(tight-binding like) obtained from DFT calculations describing a monolayer 
of 2H-NbSe$_2$. 
 
The spin susceptibility, that governs the non-local spin response to magnetic 
perturbations, is given by 
 \begin{equation}
 \chi_{ab}^{\eta\eta'}(\vec{r},\vec{r}',t)=-i\theta(t)\left\langle \left[ S_a^\eta(\vec{r},t),S_b^{\eta'}(\vec{r}',0)\right]\right\rangle,
 \end{equation}
where $a,b=x,y,z$ label the spin channel, and $\eta,\eta'$ label the atomic orbitals 
inside the unit cell. In the frequency-momentum domain we have
 \begin{equation}
 \chi_{ab}^{\eta\eta'}(\vec{q},\omega)= \int_{-\infty}^\infty dt e^{i\omega t}\int d\vec{r}
e^{i\vec{q}\cdot\vec{r}}\chi_{ab}^{\eta\eta'}(\vec{r},0,t)
 \end{equation}
In the following we compute the spin-response in the RPA approximation,
\begin{equation}
	\chi=\left[1-U\chi_0\right]^{-1}\chi_0
\label{RPA}
\end{equation}
where $\chi_0$ is the  non-interacting 
($U=0$) spin susceptibility tensor, for which closed analytical expressions are readily obtained in terms 
of the single-particle states and energies.  Therefore, equation (\ref{RPA}) permits
one to obtain the spin response including the effect of the interactions in the 
RPA\cite{antc:2010:SOCMethod,antc:2020:CrI3}. For systems with spin rotational 
invariance, such as paramagnets without spin-orbit coupling, the spin response 
matrix is proportional to the unit matrix in the spin index. Therefore, the spin response 
is the same in all directions. Here we study the case where spin rotational invariance 
is broken in the paramagnetic phase, due to SOC. 
 
We now apply this formalism to an extended Kane-Mele Hubbard model on a bipartite
honeycomb 
lattice\cite{rachel2010,soriano2010,fukaya2016,wu2019}. This is a toy model for a 
TMD. We assume that the $A$ triangular sublattice of the 
honeycomb hosts the Nb atom, whereas the $B$ sublattice contains a non-interacting site. 
The Hamiltonian is  given by
\begin{equation}
 H=H_0 + H_{SOC} + \frac{\Delta}{2}\sum_{i\sigma}\tau_i^z c^\dagger_{i\sigma}c_{i\sigma} + U\sum_{i\in A}  n_{i\uparrow}n_{i\downarrow}
\end{equation}
where $H_0$ describes first and second neighbors hopping in a honeycomb lattice,  
$H_{SOC}$ is the Kane-Mele SOC\cite{Kane2005},
\begin{equation}
\label{KMHamiltonian}
H_{SOC} =it_{\mathrm{KM}}\sum_{\langle\langle i,j \rangle\rangle,\sigma}\sigma  c_{i,\sigma}^\dagger  \hat{z}\cdot(\mathbf{d}_{kj} \times \mathbf{d}_{ik}) c_{j,\sigma}  
\end{equation}
where $\langle\langle i,j \rangle\rangle$ denotes a sum over all pairs of second neighbors $i,j$, 
 in the honeycomb lattice, and  $\mathbf{d}_{kj}$ ( $\mathbf{d}_{ik}$) are the unit vectors going from site $k$ ($i$) to site  $j$ ($k$), where   $k$ labels the common first neighbor of sites $i$ and $j$\cite{Kane2005}. The main role of $\hat{z}\cdot(\mathbf{d}_{kj} \times \mathbf{d}_{ik})$ is to make the SOC term odd under spatial inversion and with opposite sign at each sublattice, for a given direction. 

If we take $U=\Delta=0$ this term opens up a topological gap at the Dirac point.    
However, here we include a sublattice potential,  $\frac{\Delta}{2}\tau_i^z$, where 
 $\tau_i^z=\pm 1$ for $i=A,B$.~\cite{Kane2005} This breaks inversion symmetry, 
opens up a trivial gap when $\Delta\gg t_{KM}$ and, combined with the SOC term,  
leads to a spin splitting of the bands, as described above. This makes our model 
different from the case with inversion symmetry\cite{fukaya2016,wu2019}.
Since fluctuating moments are expected to be hosted by the Nb atoms, we consider 
a model where Hubbard $U$ interactions are only active in one sublattice. 

In the non-interacting limit ($U=0$), the energy bands of the Hamiltonian 
capture the main features of TMD monolayers: 
a gap separates a valence and a conduction band whose extrema are at the $K$, $K'$ 
corners of the BZ zone. In the neighborhood of the $K,K'$ points the bands have 
large spin-splitting and Kramers partners have opposite wave vectors. In this 
region the bands are well described by a Dirac equation with a mass \cite{Xiao12}.  
The model conserves the spin projection perpendicular to the atomic plane, so that 
we can still label the single-particle states 
with $\sigma=\pm 1/2$.     

\begin{figure}
\includegraphics[width=\columnwidth]{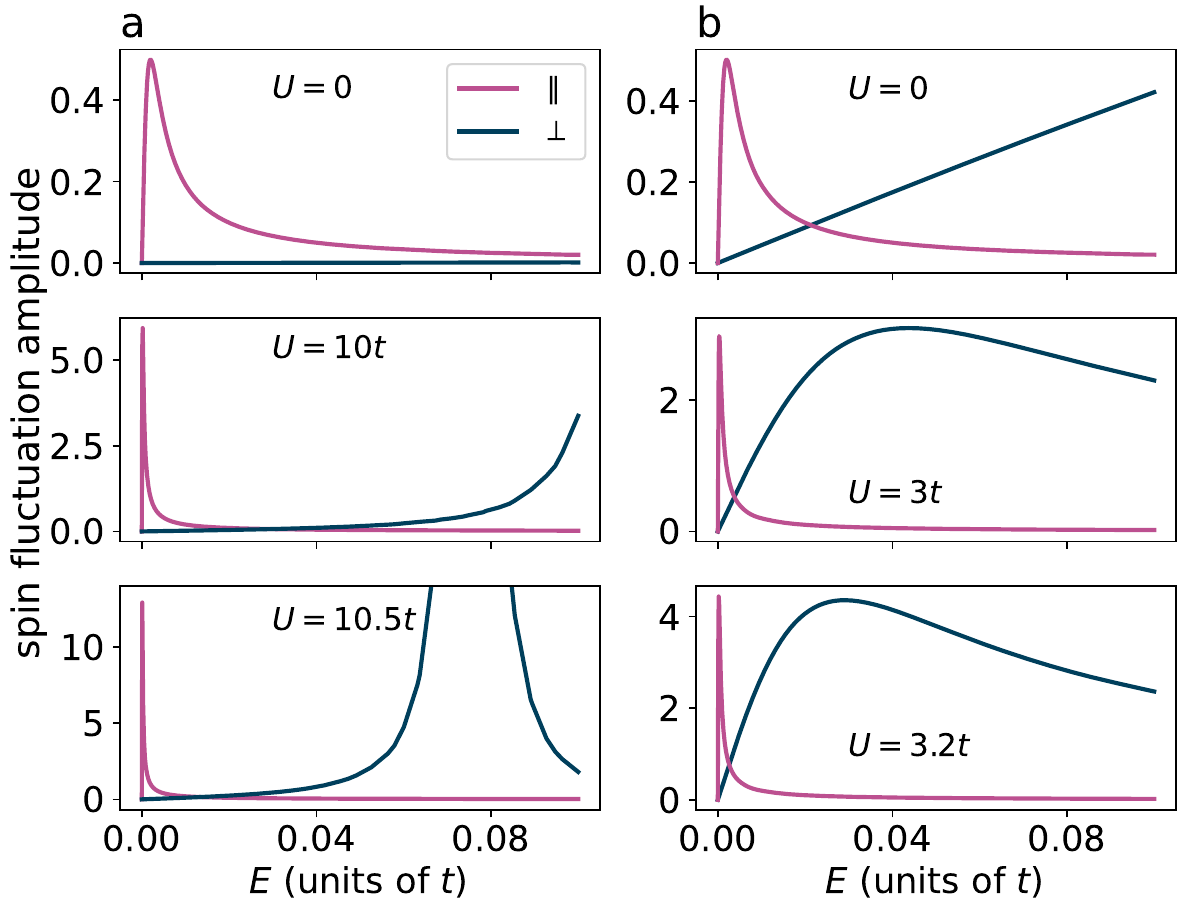}
\caption {Spin fluctuation amplitudes at zero wave vector for two Fermi energies: 
a) $E_F=3t$ (crossing only spin-split bands), and b) $E_F=1.8t$ (crossing degenarate
spin bands). In both cases the SOC strength
is $t_{KM}=0.04t$. The top row shows the mean-field spin fluctuations. The 
remaining rows show how the RPA spin fluctuation spectra change as the 
interaction strength $U$ approaches the critical value. Purple lines correspond to longitudinal 
fluctuations ($-\Im\chi^\parallel/\rho_0$) and dark blue lines correspond 
to transverse fluctuations ($-\Im\chi^\perp/\rho_0$), where $\rho_0$ is the
density of electronic states at the Fermi level, $E_F$.}
\label{fig2}
\end{figure}

Depending on the location of the Fermi energy, the model can mimic a semiconducting 
TMD, such as 2H-MoS$_2$, 2H-MoSe$_2$, 2H-WS$_2$, 
2H-WSe$_2$, doped with either electrons or holes and the Fermi energy close to the 
band extrema, or 2H-NbSe$_2$, with the Fermi energy deep down closer to the conduction
band's minima. We now study the spin fluctuations in these two limits as a function 
of the Hubbard interaction  $U$.

We focus first on the $q=0$ low energy spin fluctuations, that govern the long wavelength 
spin response of the material. Because of the $C_3$ 
symmetry of the honeycomb lattice we have $\chi_{xx}=\chi_{yy}=\chi_{\perp}$.  
In the non-magnetic phase we have $\chi_{xy}=\chi_{yx}=0$. Therefore, the spin response 
is diagonal in the spin index, with two different components for the $zz$ ($\chi_{||}$) 
and in-plane components. When the Fermi energy is located close to the $K,K'$ points, 
zero momenta spin-flip fluctuations are strictly forbidden, for energies smaller than 
the spin-splitting. In that limit, the Fermi surface is formed by spin-polarized pockets, 
with opposite polarization, at the $K$ and $K'$ points. 
In contrast, low energy spin conserving fluctuations are allowed. With this in mind, 
the results of figure 2a, showing a dramatically different behavior for 
$\chi_{\parallel}(q=0,E)$ and $\chi_{\perp}(q=0,E)$ can be easily understood. It is apparent 
that, as $U$ increases, the paramagnon peaks only forms on the $\parallel$ or off-plane channel, 
whereas the transverse spin response is quenched.

\begin{figure}
\begin{tabular}{cc}
    \includegraphics[width=\columnwidth]{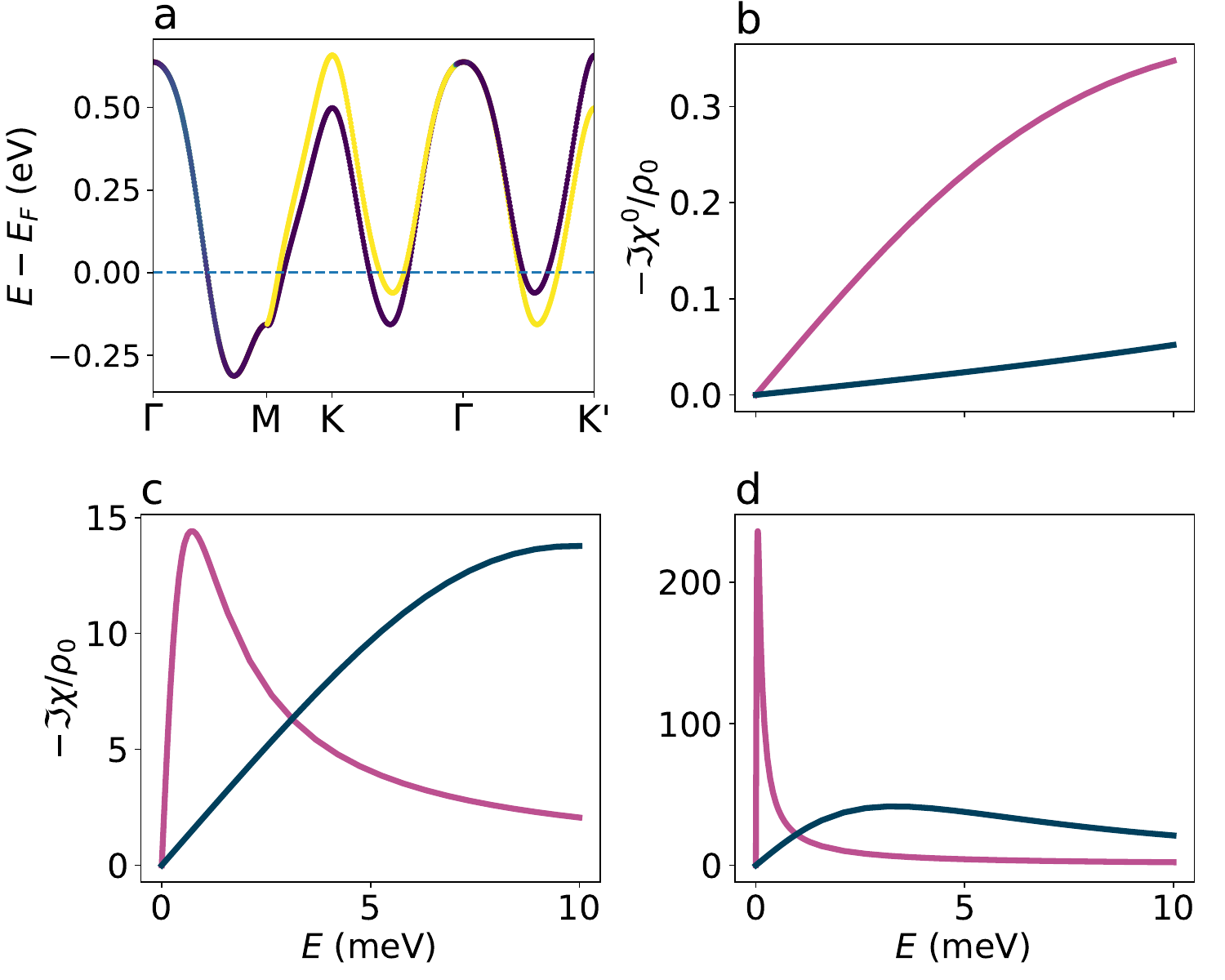} 
\end{tabular}
\caption{a) DFT bands for NbSe$_2$ around the Fermi level. The color code
represents the spin projection along $z$. 
b) Longitudinal (purple curve) and transverse (dark blue curve) mean-field spin 
fluctuation spectra at zero wave vector for NbSe$_2$, extracted from the 
DFT-based multiorbital TB model. In the lower panels we show the RPA enhanced 
spin fluctuations for c) $U=0.86$~eV and d) $U=0.89$~eV. The critical value for 
the interaction strength in this case is $U_c=0.9$~eV.  }
\label{fig3}
\end{figure}

We now address the question of whether the strong anisotropy of the spin response is 
something specific of the states close to the $K$ and $K'$ points, or, on the contrary,  
the anisotropy also occurs when the Fermi surface has spin-split bands in low symmetry 
regions of the Brillouin zone. For that matter, we consider now the case where the Fermi 
energy is located at higher in the valence band ($E_F$ corresponding to the dot-dashed line
in figure 1c). In this case we find a smaller value of the critical Stoner parameter 
$(U\rho_0)_c= 3.3$, that we attribute to a larger density of states. We 
find the same colossal anisotropy of the low-energy spin fluctuations
We refer to these very anisotropic collective modes as Ising paramagnons. 
In this case, however, the transverse fluctuations are not as strongly 
quenched as when $E_F$ only crosses spin-split bands.

We have verified that the anisotropy is driven by the combination of SOC and 
inversion symmetry breaking. For that matter we have computed the spin response for $\Delta=0$ 
and $t_{KM}>0$.  We find that the spin fluctuation spectra along the longitudinal and transverse
directions have the same lineshape and virtually identical amplitudes. 

We now ask whether the same phenomenon holds true for a more realistic Hamiltonian describing 
2H-NbSe$_2$. First we carry out density functional theory (DFT) calculations for the 2H-NbSe$_{2}$
 ~\cite{DFT1,DFT2}, using the \textsc{Quantum Espresso} suite~\cite{QE-2017}. 
 The electronic interaction  was described within the generalized gradient approximation (GGA) via the 
 Perdew-Burke-Ernzerhof (PBE) functional~\cite{pbe}. Ionic potentials were described by projector 
 augmented-wave (PAW)~\cite{PAW} pseudopotentials available in the 1.0 pslibrary database~\cite{pslibrary}. 
The wavefunctions and charge density cutoff energies were 71.5 and 715 Ry, respectively.  Full structural optimization
 was performed until Hellman--Feynman forces were smaller than 0.01 eV/\AA \ with a 13$\times$13$\times$1
  reciprocal space sampling. We found a lattice parameter of 3.47 \AA, which is in agreement with other DFT 
  calculations~\cite{NbSe2-1}. The Hamiltonian was constructed with a larger $K$-sampling of 
  27$\times$27$\times$1. We allow for spin polarization but the system converged to a non-magnetic
   ground state. Our results are in line with those obtained in the literature~\cite{NbSe2-1}

After the structural optimization, a local effective Hamiltonian was constructed via 
the pseudo-atomic orbital (PAO) projection method~\cite{PAO1,PAO3} as implemented in 
the \textsc{paoflow} code~\cite{PAO5}. The method consists in projecting the plane wave 
Kohn-Sham states onto a compact subspace spanned by PAOs already built in the PAW potentials. 
This procedure reduces the basis set from several thousand plane waves to a few atomic 
orbital-like basis functions with accuracy comparable to DFT calculations. 
In the supplementary material we compare the \textsc{paoflow} and \textsc{Quantum Espresso} 
band structures. The PAW potential for Nb and Se were constructed with a \textit{sspd} and \textit{spd} PAO basis, 
respectively. This choice results in 13 and 9 orbitals per Nb and Se atom. Obviously, 
spin-orbit-coupling is essential for spin-splitting at $K$ and $K'$ points. 
Therefore, we include it as a local term of the form
\begin{equation}
H_\mathrm{SOC}=\sum_{l}\sum_{\mu\nu}\sum_{\sigma,\sigma'=\uparrow,\downarrow}
\xi_l(\vec{L}\cdot\vec{S})_{l\mu\sigma,l\nu\sigma'} a^\dagger_{l\mu\sigma} a_{l\nu\sigma'}  ,
\end{equation}
where $l$ is an atomic site index, $\mu,\nu$ are orbital indices,
and $\sigma,\sigma'$ are spin indices. $\vec{L}$
is the orbital angular momentum operator and $S$ is the electronic 
spin operator. The orbital indices $\mu,\nu$ run over the $p$ orbitals 
when atomic site $l$ is occupied by a Se atom, and over the $d$ orbitals
when $l$ is occupied by a Nb atom. The SOC intensities at Se and Nb atoms
have been adjusted such that the multiorbital LCAO model with local SOC
reproduces as faithfully as possible the energy bands resulting from a fully 
relativistic DFT calculation. We find that the best fit is given by 
$\xi_\mathrm{Nb}=79$~meV and $\xi_\mathrm{Se}=211$~meV, in line with
those reported in reference~\onlinecite{Kosmider13b}. Explicit comparison
between the LCAO and the DFT bands is given in the supplementary material.

We now apply the RPA method for our multi-orbital tight-binding model. The on-site 
atomic  Coulomb repulsion interaction is given by the Hamiltonian:
\begin{equation}
H= \sum_l U_l\sum_{\mu\nu}\sum_{\sigma\sigma'} a^\dagger_{l\mu\sigma} a^\dagger_{l\nu\sigma'}
a_{\nu\sigma'}a_{\mu\sigma}
\end{equation}
where $U_l$ is taken as a free parameter in the calculations, 
as we have done in the case of the KM Hubbard model. When atomic site $l$ is occupied by
a Se atom we take $U_l=0$. $\mu,\nu$ are orbital indices running over the $d$ 
orbitals centered on the Nb sites, and $\sigma_1,\sigma_2$ are spin indices. We find that $U_c= 0.9$~eV  
marks the critical value of the instability to a ferromagnetic phase.

The  $q=0$ spin susceptibility matrix,  calculated within the RPA,  using the  multi-orbital DFT based TB model are shown in 
figure \ref{fig3}.   We find again a very anisotropic response, with paramagnon 
enhancement in the $\perp$ channel, much larger than the in-plane spin fluctuations. We also 
find some quantitative differences. For instance, spin-flip fluctuations are not completely 
quenched at small energy, in contrast to the KM model. We attribute this difference to the 
fact that in the multi-orbital DFT based TB model $S_z$ is no longer a conserved quantity and 
the states away from the $K,K'$ points have a non-negligible mixing of the $\uparrow$ 
and $\downarrow$ channel. Yet, the main result of this work, the large anisotropy of the spin fluctuations remains. 

Some degree of control over the effective Coulomb interaction strength $U$ 
is available through, for instance, the effective dielectric constant of a
conveniently chosen substrate~\cite{Raja2017,vanloon2020coulomb}. Its effects
can also be controlled indirectly changing the degree of electronic confinement,
either through strain applied to large-area samples~\cite{Yan2020} or by producing 
samples of smaller sizes~\cite{Ganguli2022}.

\begin{figure}
\begin{tabular}{cc}
    \includegraphics[width=\columnwidth]{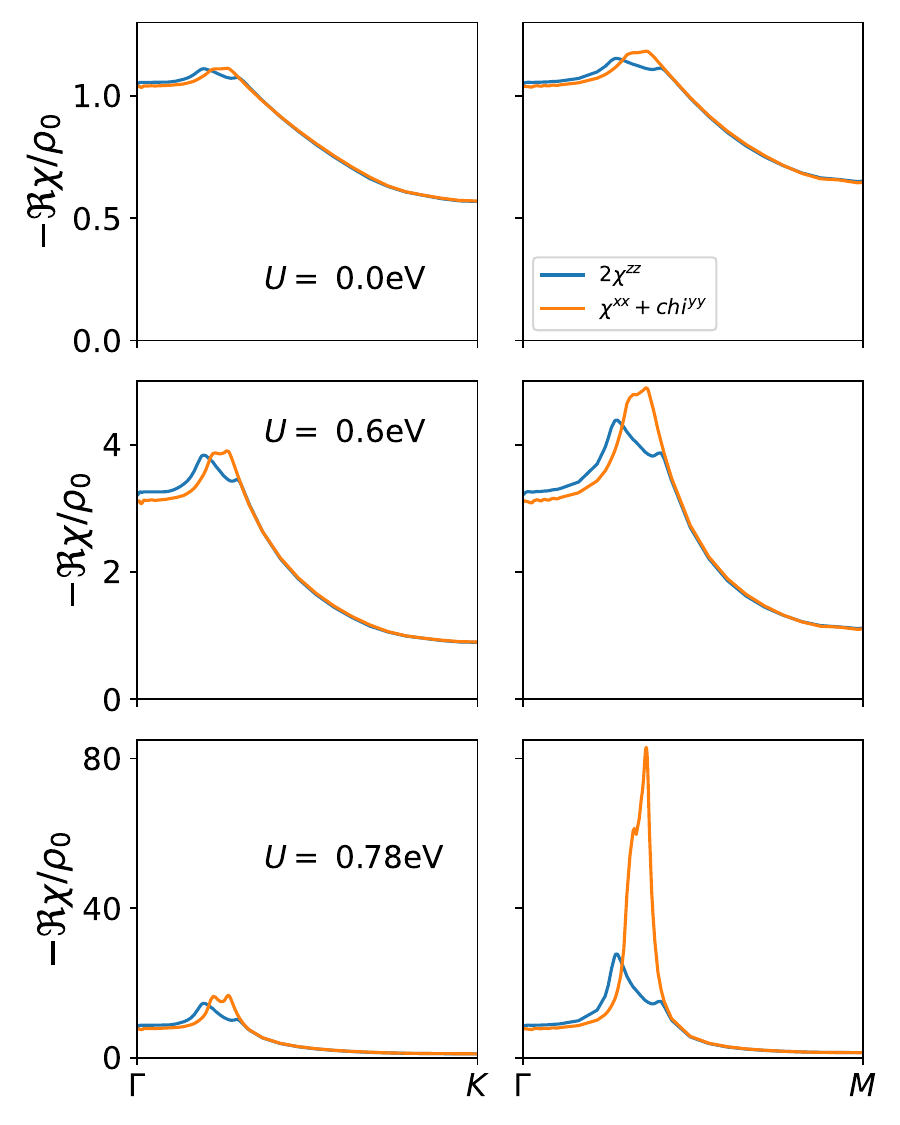} 
\end{tabular}
\caption{Transverse ($\chi^{xx}+\chi^{yy}$) and longitudinal ($\chi^{zz}$) 
spin susceptibilities for 2H-NbSe$_2$ as functions
of wave vector along the $\Gamma-K$ (left panels) and $\Gamma-M$ (rigth panels) lines, 
in the static limit ($E=0$). Top panels are the mean-field results ($U=0$), the remaining panels show 
RPA results for different values of the interaction strength $U$.}
\label{fig:Rechistatic}
\end{figure}

We now turn to the finite wave vector 
spin susceptibility in the static limit, 
 directly related to the effective pairing interactions in spin fluctuation mediated
superconductivity\cite{Sigrist2005}. In figure~\ref{fig:Rechistatic} we show
the real part of the transverse and longitudinal spin susceptibilities 
at $E=0$ for 2H-NbSe$_2$, as a function of wave vector, for different values
of $U$. These results have been obtained using the DFT-derived fermionic
hamiltonian for 2H-NbSe$_2$. Their most prominent feature is the divergency 
of the \textsl{transverse} susceptibility  around $q\sim 0.21(2\pi/a_0)$
as $U$ is ramped up, while the longitudinal susceptibility remains finite.
Importantly, the divergency here happens for $U\sim 0.8$~eV, which is 
\textsl{significantly smaller} than the corresponding value ($\sim 0.9$~eV) 
for the uniform ($q=0$) susceptibility. This indicates that the magnetic 
instability in 2H-NbSe$_2$ is actually in-plane, of \textsl{spin density wave 
(SDW) nature}, instead of out-of-plane ferromagnetic.
 At finite frequencies, we find XY paramagnons: a strong enhancement 
of the transverse spin fluctuations while longitudinal fluctuations are only modestly
enhanced (see  figure~6 on the supplementary material).
The strongly anisotropic response and the proximity to a 
SDW instability may have implications for spin-fluctuation mediated
pairing~\cite{Sigrist2005}. Qualitatively similar behavior has been analyzed, 
for instance, in references~\onlinecite{Romer2016PRB,Romer2019PRL}. It is also
worth mentioning that the spin fluctuation spectrum of 2H-NbSe$_2$ depends strongly
on the direction of the wave vector, a feature that is relevant to the symmetry
of the pairing interactions.

Direct observation of paramagnons, either $q=0$ Ising paramagnons, 
or finite $q$ XY paramagnons,  with energy and momentum resolution is presently possible
only via neutron scattering~\cite{ParamagnonsPd2010}. However, the applicability 
of this technique is restricted to bulk samples, due to the very weak 
neutron-electron interaction (through the dipolar fields produced by their spin 
magnetic moments). An alternative would be to prepare multilayer samples of 
NbSe$_2$ separated by a non-magnetic insulator (such as hexagonal boron nitride,
for example). This would preserve the 2D character of the NbSe$_2$ paramagnons while
providing the needed cross-section for neutron scattering.

In conclusion, we have calculated the spin fluctuations of spin-valley coupled systems 
that describe non-centrosymmetric TMD, such as doped 2H-MoS$_2$ 
and 2H-NbSe$_2$ monolayers. We have used both toy model Hamiltonians, such as the 
Kane-Mele-Hubbard model, and DFT-based models. We have considered both $q=0$ and finite $q$.
 In all  cases we find a very large 
spin anisotropy of the spin response, driven by the interplay of SOC
and lack of inversion symmetry. Remarkably, the magnetic anisotropy of paramagnons is 
wave-vector dependent, so that we have Ising Paramagnons for $q=0$ are  XY paramagnons for finite $q$.
Our calculations reveal that  2H-NbSe$_2$ monolayers are closer to  
SDW instability  with in-plane easy axis, rather than a ferromagnetic ($q=0$) off-plane instability. 
Our findings can have profound implications for the nature of both the normal state and
the superconducting phase of 2H-NbSe$_2$.

We acknowledge fruitful discussions with Ilya M. Eremin. We acknowledge financial support 
from the Ministry of Science and Innovation of Spain (grant No. PID2019-109539GB-41), 
from Generalitat Valenciana (grant No. Prometeo2017/139) and from Fundação Para a Ciência e
a Tecnologia, Portugal (grant No. PTDC/FIS-MAC/2045/2021).

\clearpage

\onecolumngrid

\appendix

\section{Further details regarding the electronic structure of monolayer 2H-NbSe$_2$}

\begin{figure}
\centering
\includegraphics[width=\columnwidth]{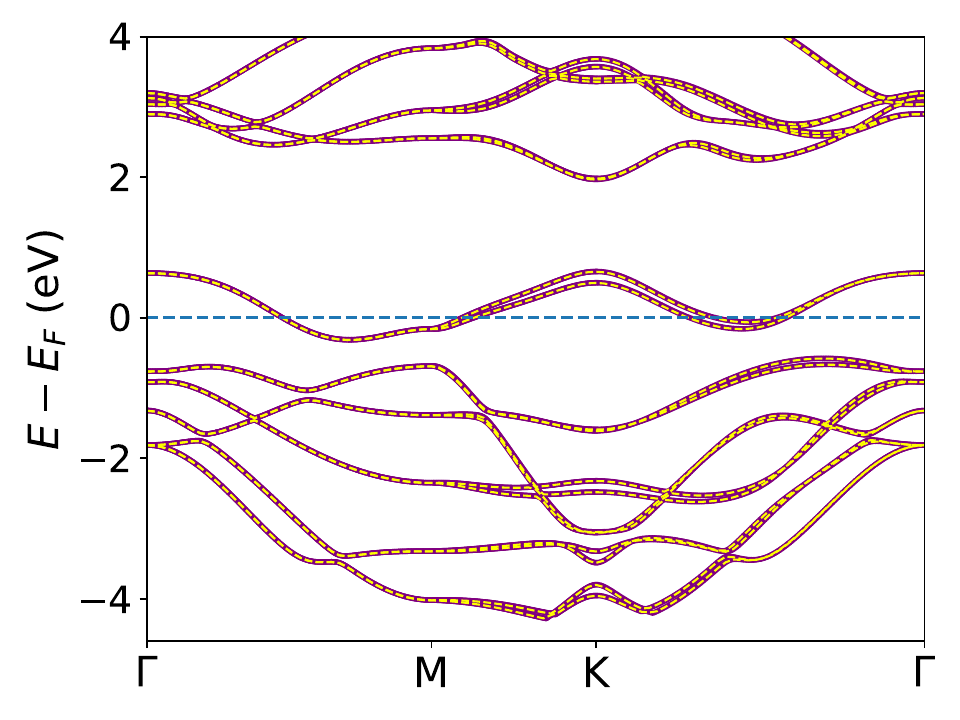}
\caption{Comparison between the band structure provided by the 
fully relativistic DFT calculation described in the main text 
(dashed yellow lines) and the energy bands generated by the
a tight-binding-like hamiltonian, including local spin-orbit 
coupling (purple symbols).}
\label{fig:compareTBDFT}
\end{figure}

Here we provide additional plots showing results for the DFT calculation 
and the associated multiorbital tight-binding model. We show in figure~\ref{fig:compareTBDFT}
that the multiorbital model derived from the DFT calculation, supplemented by a local
spin-orbit coupling term, fits exceedingly well the DFT bands.

\begin{figure}
\centering
\includegraphics[width=\columnwidth]{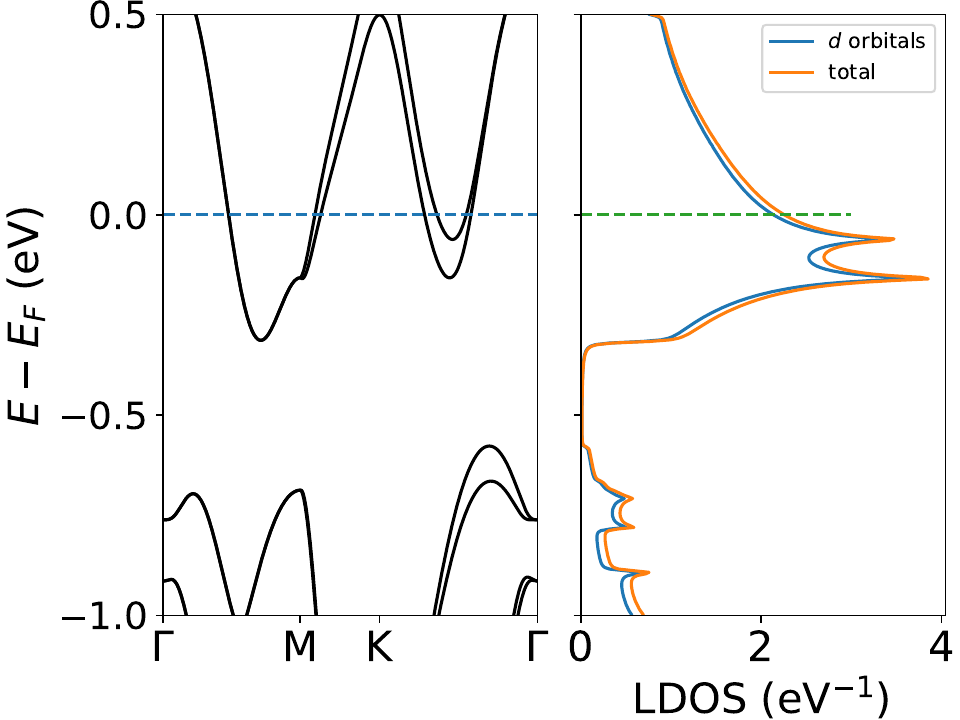}
\caption{Detail of the band structure (left) and local density of states 
at Nb sites (right) around the Fermi level $E_F$ for a NbSe$_2$ monolayer. 
We also show the LDOS projected on the $d$ orbitals (right panel, blue curve). 
These results were obtained using the PAO hamiltonian, and include SOC.}
\label{fig:LDOSmultiorb}
\end{figure}

In figure~\ref{fig:LDOSmultiorb} we show the local density of states 
around the Fermi level, projected on the Nb site. We also show how the
energy eigenstates around the Fermi level have predominantly $d$ character.
From the $d$-projected LDOS it can also be inferred that the critical
$U$ for which the spin-unpolarized ground state becomes unstable
(to a spatially uniform perturbation) is $U_c \approx 0.93$~eV. 
Notice, however,that the non-uniform transverse susceptibility 
diverges at a finite wave vector for values of $U$ that are considerably 
smaller than that.

\section{Paramagnon spectra at finite wave vectors}

\begin{figure}
\begin{tabular}{cc}
    \includegraphics[width=\columnwidth]{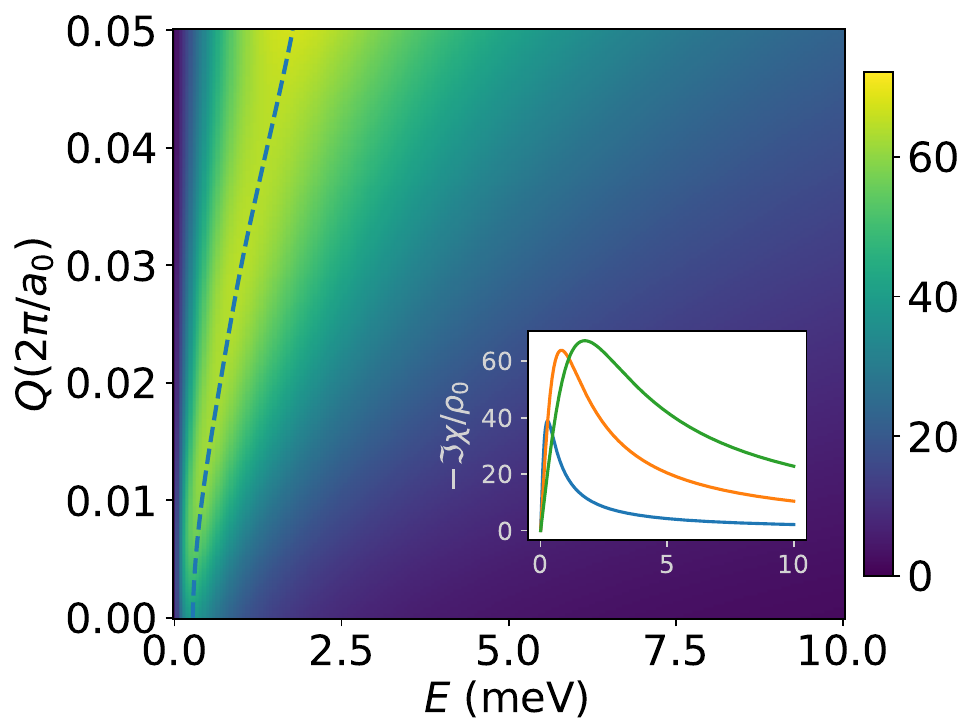} 
\end{tabular}
\caption{Density plot of the spectral density of longitudinal spin fluctuations in 
monolayer NbSe$_2$ as a function of energy and wave vector for $U=0.88$~eV. 
The dashed line marks the positions of the maxima of the spectral density. 
The inset shows the same spectral density as a function of energy for three
values of the wave vector (along $\Gamma-K$): 0 (blue curve), $0.01(2\pi/a_0)$ (orange curve) 
and $0.05(2\pi/a_0)$ (green curve). }
\label{fig4}
\end{figure}

If figure~\ref{fig4} we show the longitudinal spin spectral density as
a function of wave vector and energy. These results have been obtained
with the multi-orbital model extracted from the DFT calculation.
As wave vector increases, the energy at which the spectral density 
peaks also increases; by following this peak we extract a 
``paramagnon dispersion relation,'' which can serve as a guide for the observation
of the Ising paramagnons of NbSe$_2$ in experiments.

\begin{figure}
\begin{tabular}{cc}
    \includegraphics[width=\columnwidth]{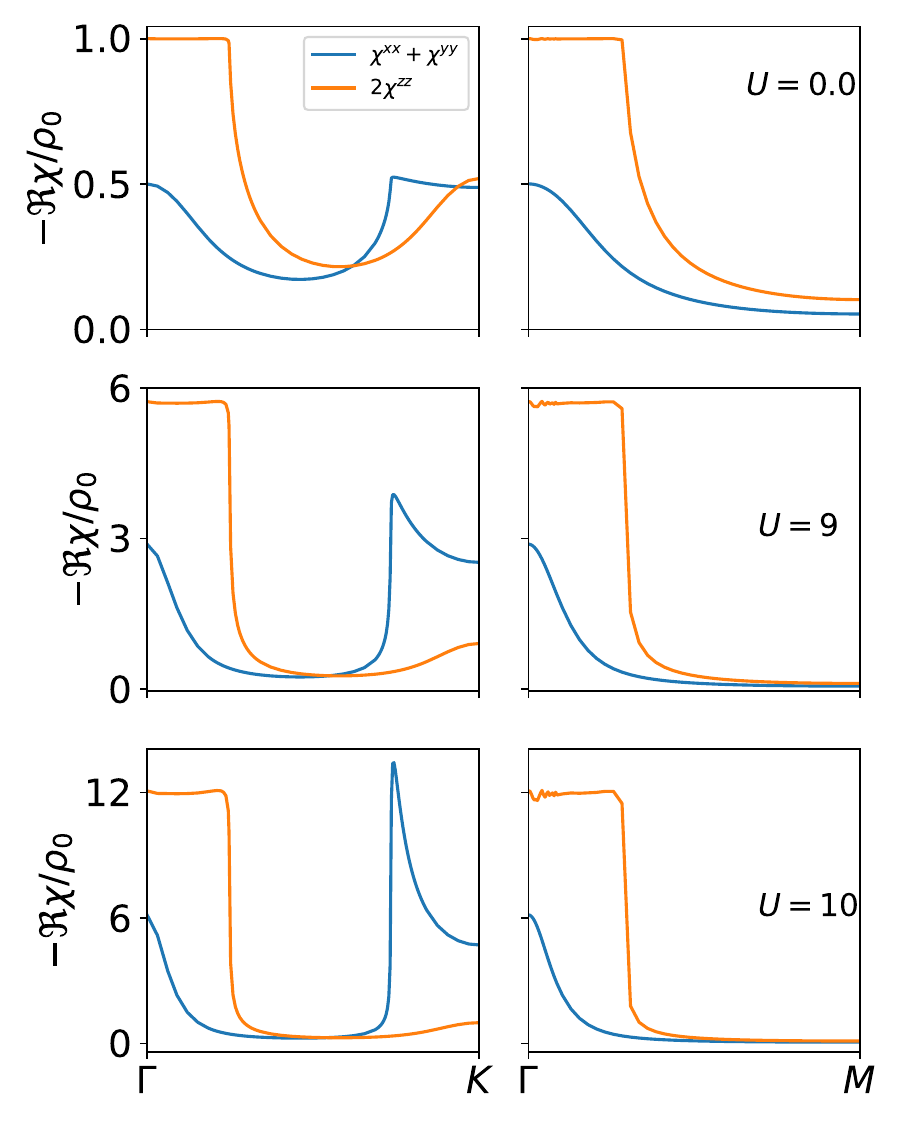} 
\end{tabular}
\caption{Transverse ($\chi^{xx}+\chi^{yy}$) and longitudinal ($\chi^{zz}$) 
spin susceptibilities for the Kane-Melle-Hubbard model as functions
of wave vector along the $\Gamma-K$ (left panels) and $\Gamma-M$ (rigth panels) lines, 
in the static limit ($E=0$). Top panels are the mean-field results ($U=0$), the remaining panels show 
RPA results for different values of the interaction strength $U$.}
\label{fig:RechistaticKM}
\end{figure}

\begin{figure}
\begin{tabular}{cc}
    \includegraphics[width=\columnwidth]{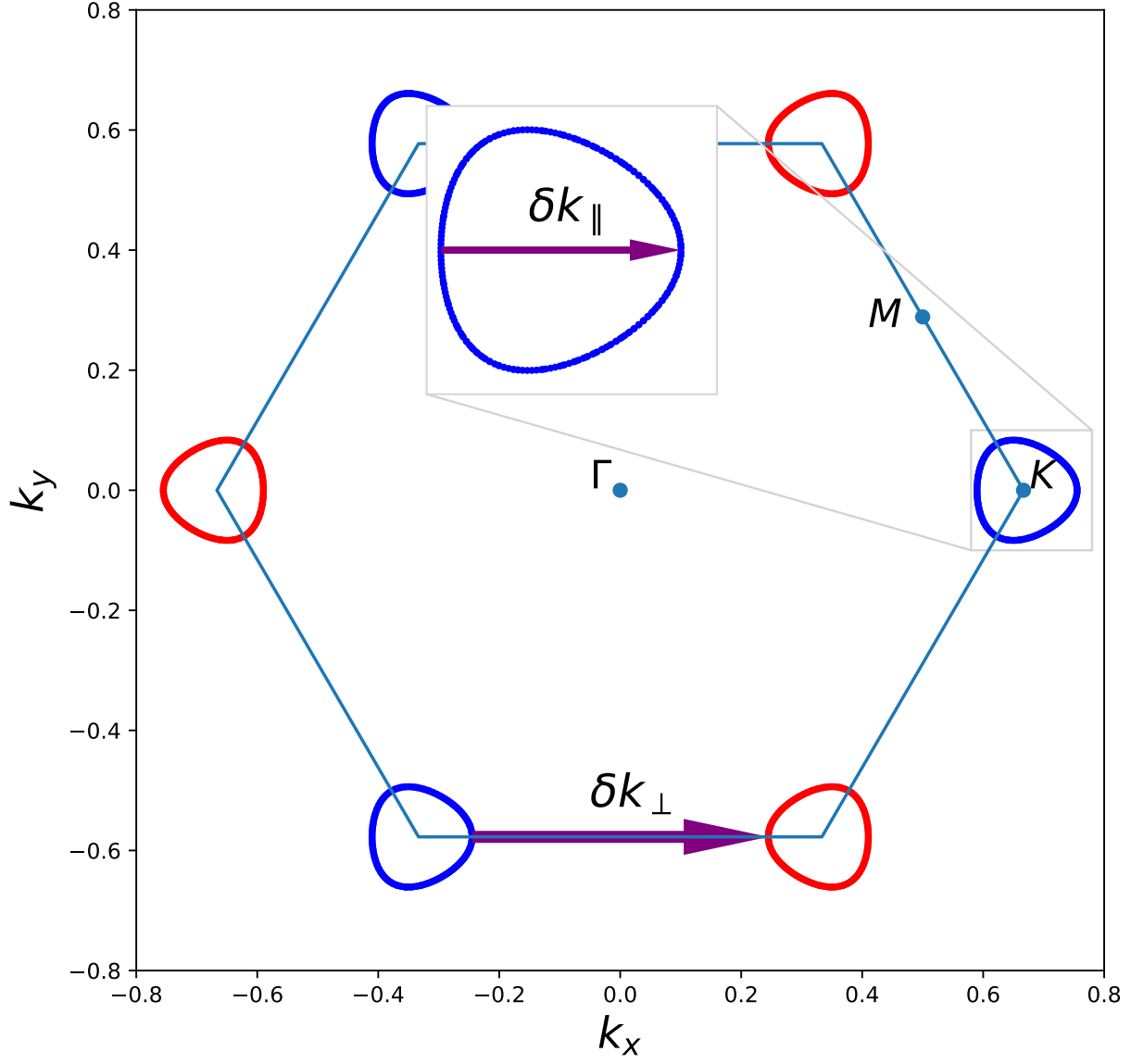} 
\end{tabular}
\caption{
Fermi surface for the Kane-Melle-Hubbard model for $E_F=3t$. The width of the pockets
	around the $K$ points ($\delta k_\parallel$) correspond to the region in which $\Re\chi^{zz}(Q,0)$
is almost flat (see figure \ref{fig:Rechistatic}). The length of the nesting
	vector connecting opposite-spin pockets around $K$ and $K'$ ($\delta k_\perp$) corresponds to the
position of the peak in $\chi^{xx}+\chi^{yy}$ along $\Gamma-K$.}
\label{fig:FStriang}
\end{figure}

In figure~\ref{fig:RechistaticKM} we show the transverse and longitudinal 
components of the spin susceptibility for the toy model, at zero
excitation energy, as a function of wave vector. The spin-valley
locking leads to large differences between the two responses. The
features of these response functions can be associated to nesting
vectors connecting different portions of the Fermi surface (FS),
as shown in figure~\ref{fig:FStriang}. The vector connecting
parallel portions of the FS with the same spin polarization,
$\delta k_\parallel$, is associated with the region over which 
the longitudinal mean-field ($U=0$) susceptibility $\chi^{zz}$ 
is flat. The kink seen in the transverse component ($\chi^{xx}+\chi^{yy}$)
at a finite wave vector along $\Gamma-K$ is associated with the
wave vector connecting portions of the FS with opposite spins,
$\delta k_\perp$.

\begin{figure}
\begin{tabular}{cc}
    \includegraphics[width=\columnwidth]{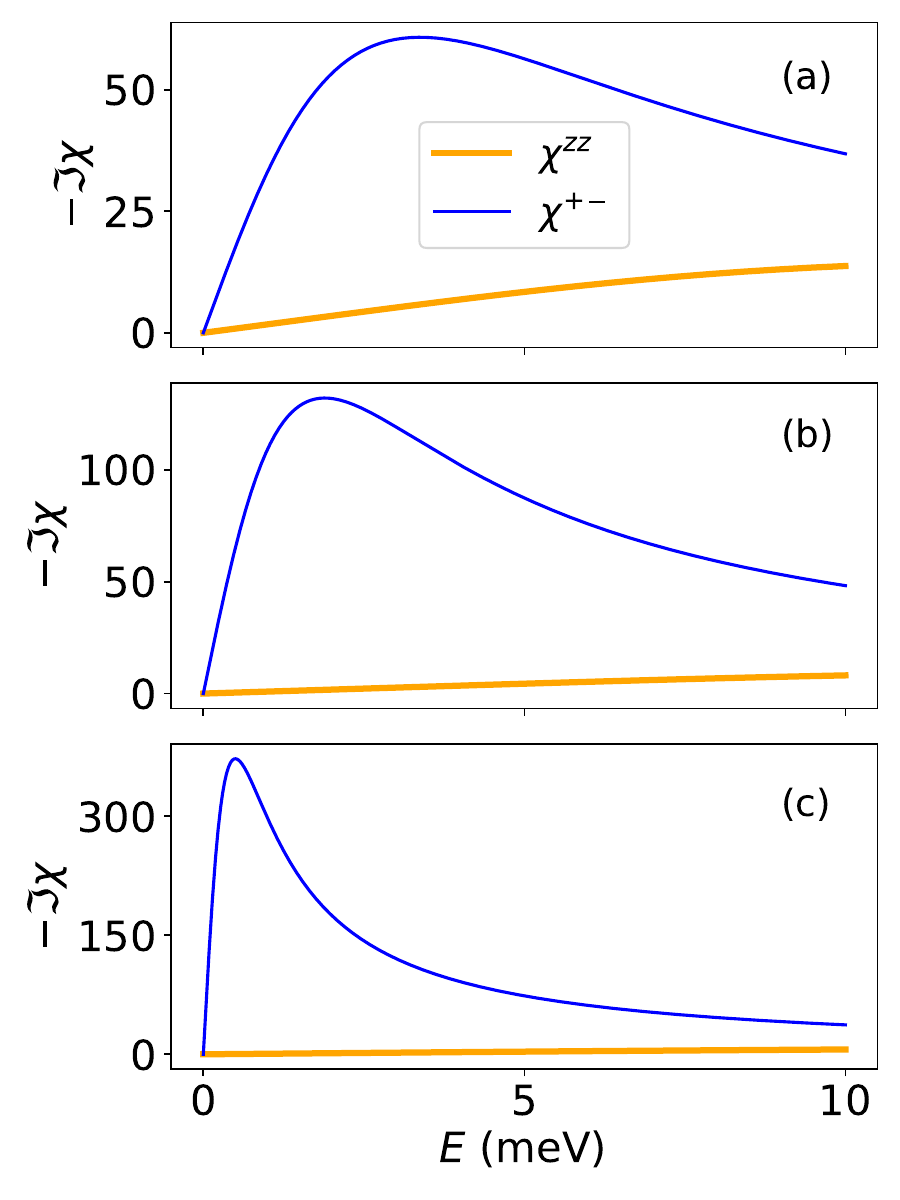} 
\end{tabular}
\caption{Transverse ($\chi^{+-}$, blue curve) and longitudinal ($\chi^{zz}$, orange)
spin fluctuation spectral densities at $q=0.18(2\pi/a_0)$ (a),  $q=0.20(2\pi/a_0)$ (b) and $q=0.21(2\pi/a_0)$ (c), along the $\Gamma-\mathrm{M}$ 
line, for $U=0.79$~eV.}
\label{fig:XYpmag}
\end{figure}

In figure~\ref{fig:XYpmag} we show the transverse and longitudinal spin fluctuation spectra
at three different wave vectors along the $\Gamma-\mathrm{M}$ line, for $U=0.79$~eV. Besides being
strongly enhanced, transverse fluctuations clearly dominate the spectrum in this region of
the Brillouin zone, prompting us to identify the existence of XY paramagnons.

\end{document}